\documentclass[10pt,twocolumn]{article}
\usepackage[margin=0.85in]{geometry}
\usepackage[T1]{fontenc}
\usepackage[utf8]{inputenc}
\usepackage{booktabs}
\usepackage{amsmath,amssymb}
\usepackage{graphicx}
\usepackage[hypertexnames=false]{hyperref}
\usepackage{array}
\usepackage{float}
\usepackage{xcolor}
\usepackage{natbib}
\usepackage{microtype}
\usepackage{caption}
\title{What Topological and Geometric Structure Do Biological\\Foundation Models Learn? Evidence from 141 Hypotheses}
\author{%
  Ihor Kendiukhov\\
  \small Department of Computer Science\\
  \small University of T\"ubingen\\
  \small T\"ubingen, Germany\\
  \small \texttt{kenduhov.ig@gmail.com}
}
\date{}

\begin{document}
\maketitle

\begin{abstract}
When biological foundation models like scGPT and Geneformer learn to process single-cell gene expression, what kind of geometric and topological structure forms in their internal representations?
Is that structure biologically meaningful, or merely an artifact of training?
And how confident should we be in such claims?

We address these questions through autonomous large-scale hypothesis screening: an AI-driven executor--brainstormer loop that proposed, tested, and refined 141 geometric and topological hypotheses across 52 iterations, covering persistent homology, manifold distances, cross-model alignment, community structure, directed topology, and more---all with explicit null controls and disjoint gene-pool splits.

Three principal findings emerge.
First, \textbf{the models learn genuine geometric structure}: gene embedding neighborhoods exhibit non-trivial topology (persistent homology significant in 11/12 transformer layers at $p < 0.05$ even in the weakest domain, and 12/12 in the other two), a multi-level distance hierarchy where manifold-aware metrics outperform Euclidean distance for identifying regulatory gene pairs, and graph-community partitions that track known transcription factor--target relationships.
Second, \textbf{this structure is shared across independently trained models}: CCA alignment between scGPT and Geneformer yields canonical correlation of 0.80 and gene retrieval accuracy of 72\%---yet no method among 19 tested could reliably recover gene-level correspondences (the sole partial exception, perturbation-response alignment, failed to replicate across seeds), revealing that the models agree on the ``shape'' of gene space but not on precise gene placement.
Third, \textbf{the structure is more localized than it first appears}: under the most stringent null controls (simultaneous auditing against all null families), robust signal concentrates in immune tissue, while lung and external-lung signals become fragile.

These results---especially the carefully documented negatives among 141 hypotheses---calibrate what we can and cannot extract from biological model geometry, and demonstrate how autonomous screening can efficiently map the boundary between real structure and statistical artifact.
\end{abstract}

\section{Introduction}

Foundation models for single-cell genomics have demonstrated impressive performance on downstream tasks, from cell-type annotation to gene perturbation prediction~\citep{cui2024scgpt,theodoris2023transfer,yang2022scbert,hao2024large}.
But an unsettling question lingers beneath these successes: \emph{what do these models actually understand about biology?}
Are their internal representations organized in biologically meaningful ways, or are they opaque statistical summaries that happen to correlate with biological outcomes?

The stakes are high.
If these models encode meaningful biological structure internally---regulatory relationships, protein interactions, pathway organization---then their representations could be mined for biological discovery~\citep{lopez2018deep,luecken2019current}.
If not, claims about ``biological knowledge'' in foundation models rest on fragile ground.

Mechanistic interpretability research in language models has shown that neural representations can be surprisingly structured, encoding linear concept representations~\citep{park2023linear}, feature superposition~\citep{elhage2022toy}, depth-dependent processing stages~\citep{nanda2023progress}, and recoverable computational circuits~\citep{conmy2023towards,olah2020zoom}.
In the biological domain, recent work has examined this question through two complementary lenses.
First, analysis of attention patterns in scGPT and Geneformer revealed that attention encodes protein-protein interactions in early layers and transcriptional regulation in later layers---but provides \emph{no incremental value} for perturbation prediction beyond trivial gene-level baselines~\citep{kendiukhov2025attention}.
Second, spectral analysis of residual-stream geometry showed that scGPT organizes genes along interpretable axes corresponding to subcellular localization, interaction networks, and regulatory identity~\citep{kendiukhov2026spectral}.

These studies established that biological transformers encode \emph{linear} geometric structure.
But a fundamental question remains: does the embedding space contain deeper \emph{nonlinear} geometric and topological structure---loops, manifold curvature, community boundaries, directional asymmetries---that carries biological meaning beyond what linear analysis captures~\citep{carlsson2009topology,rabadan2019topological}?
And critically, how much of any apparent structure survives rigorous null controls?

Traditional hypothesis-driven approaches are poorly suited to this question because the space of possible geometric and topological properties is vast, and the risk of reporting only positive findings creates publication bias.
We therefore adopt a fundamentally different approach: \textbf{autonomous hypothesis screening}.
An AI-driven executor--brainstormer loop autonomously proposed, executed, and evaluated 141 distinct hypotheses across 52 iterations, systematically recording both positive and negative results.
Each hypothesis was tested against explicit null models, with results replicated across three tissue domains, three random seeds, and disjoint gene-pool splits designed to prevent information leakage.

What emerges is a nuanced picture---neither the triumphant narrative that ``the models encode rich topological biology'' nor the nihilistic conclusion that ``it's all artifacts.''
The truth lies in between, and the location of that boundary is precisely what this paper maps.

\section{Methods}

\subsection{Autonomous Hypothesis Screening}

The research was conducted by an autonomous executor--brainstormer loop powered by a large language model (OpenAI Codex 5.3, ``extra high'' reasoning setting).
In each iteration, the \textbf{executor} receives a hypothesis specification, writes and runs a self-contained Python experiment on pre-extracted foundation model embeddings, and produces a quantitative report with effect sizes, null-calibrated $p$-values, and pass/fail verdicts.
The \textbf{brainstormer} reviews cumulative results from all prior iterations, identifies under-explored regions of the hypothesis space, and proposes 2--4 new hypotheses for the next iteration, explicitly informed by prior negative results to avoid redundant testing.

This design serves two purposes.
First, the brainstormer steers the search adaptively, building on successful branches (e.g., the triangle-defect spectrum led to motif--community hypotheses) while retiring failed lineages early.
Second, by recording every negative and inconclusive result alongside positive ones, the system builds a comprehensive map of what the embedding space does and does not encode---something that traditional research pipelines, optimized for positive findings, systematically fail to produce.

The loop ran for 52 productive iterations (53 total; one initialization failure), testing 141 hypotheses organized into nine families (Table~\ref{tab:families}).

\begin{table}[t]
\centering
\caption{Hypothesis families and their biological motivation.}
\label{tab:families}
\small
\begin{tabular}{@{}p{2.8cm}p{4.4cm}@{}}
\toprule
Family & Biological question \\
\midrule
Persistent homology & Are there ``loops'' in gene-embedding space reflecting cyclic regulatory motifs? \\
\addlinespace
Manifold distances & Do curved manifold distances better capture regulatory proximity than straight-line distance? \\
\addlinespace
Cross-model alignment & Do independently trained models converge on the same geometric organization of genes? \\
\addlinespace
Community structure & Do embedding neighborhoods form clusters that correspond to regulatory modules? \\
\addlinespace
Directed topology & Does the geometry encode directional information (TF $\to$ target)? \\
\addlinespace
Intrinsic dimensionality & Does the local complexity of the manifold correlate with regulatory architecture? \\
\addlinespace
Sparse descriptors & Can combining multiple geometric features through stability selection outperform any single metric? \\
\addlinespace
Motif--community & Do signed regulatory motifs (activation/repression) align with geometric community structure? \\
\addlinespace
Other geometric & Do additional geometric properties (curvature, hyperbolicity, anisotropy) carry regulatory information? \\
\bottomrule
\end{tabular}
\end{table}

\subsection{Models and Data}

We analyzed \textbf{scGPT}~\citep{cui2024scgpt} (12 transformer layers~\citep{vaswani2017attention}) by extracting residual-stream gene embeddings: for each tissue context, single-cell expression profiles from the Tabula Sapiens atlas~\citep{tabulasapiens2022} were fed through the pretrained model, and the hidden-state vectors at each transformer layer were averaged across cells to produce a single embedding vector per gene per layer.
Three tissue domains were used: lung, immune, and external-lung (a held-out lung dataset not used for any parameter tuning), each with three independent random seeds controlling gene sampling and train/test partitioning.
For cross-model comparisons, we extracted analogous gene embeddings from \textbf{Geneformer V2-316M}~\citep{theodoris2023transfer} (18 layers, 18 heads) on the same tissue domains, restricting to genes present in both models' vocabularies (typically 320 shared genes per domain).
Typical experiments sampled 160--350 genes per test, with PCA dimensionality reduction (10--20 components) applied to the embedding vectors for computational tractability and to mitigate the curse of dimensionality in distance computations.

Regulatory ground truth was drawn from DoRothEA~\citep{garcia2019benchmark} (confidence-tiered TF--target annotations), TRRUST~\citep{han2018trrust} (signed regulatory edges indicating activation or repression), STRING~\citep{szklarczyk2019string} (protein--protein interaction confidence scores), and Gene Ontology~\citep{ashburner2000gene,gene2021gene} (functional co-membership annotations).
Gene regulatory network inference from single-cell data remains a major benchmark challenge~\citep{pratapa2020benchmarking,dibaeinia2020sergio}, and the evaluation design adopted here reflects lessons from the benchmarking literature on the sensitivity of method rankings to protocol choices~\citep{saelens2019comparison}.

\subsection{Null Models}

The choice of null model turned out to be one of the most consequential decisions in this work.
We employed a hierarchy of increasingly stringent controls:

\textbf{Feature-shuffle null}: randomly permute embedding features independently across genes (20--24 replicates per condition), destroying geometric structure while preserving per-gene marginal feature distributions.
This is the weakest control, analogous to a shuffled-label baseline.

\textbf{Label-permutation null}: randomly permute regulatory edge labels (120--200 replicates), controlling for the base rate of positive edges in the evaluation set.

\textbf{Degree-preserving rewiring null}: rewire the $k$-nearest-neighbor graph ($k$ typically 12, with adaptive selection from 10--35 in some experiments) while preserving each node's degree, testing whether apparent structure is merely a byproduct of connectivity patterns rather than specific neighborhood identity.

\textbf{Coexpression-matched null}: compute pairwise absolute Pearson correlations of gene expression across cells, bin gene pairs jointly by coexpression level and graph degree (5 bins each), and permute edge labels only within each combined stratum.
This controls for the strongest known confound: the tendency for coexpressed genes to be geometrically proximal regardless of regulatory relationship.

\textbf{Strict max-null audit}: compare the observed statistic against the \emph{maximum} of the 95th percentile across \emph{all} null families simultaneously---the most conservative possible threshold.

As we will show, many signals that look convincing under weaker controls evaporate under stronger ones, making this hierarchy essential for honest interpretation.

\subsection{Evaluation Design}

All experiments used a \textbf{disjoint gene-pool split} design to prevent information leakage.
In the \emph{source-disjoint} regime, the transcription factors (regulatory sources) in the test set are entirely absent from the training set, so the model must generalize to unseen regulators.
In the \emph{target-disjoint} regime, the regulated target genes are held out instead.
Together, these two regimes ensure that no gene appears on both sides of the train/test boundary.
A finding was considered robust only if it held in \emph{both} split regimes across \emph{all} tissue domains.
Key metrics were $\Delta$AUROC (improvement over baseline in discriminating regulatory from non-regulatory gene pairs), null-gap (observed $\Delta$AUROC minus the 95th percentile of the null distribution; positive means the signal exceeds what the null can explain), and domain-split pass rate (fraction of domain $\times$ split combinations with positive null-gap).

\section{Results}

\begin{figure*}[t]
\centering
\includegraphics[width=0.85\textwidth]{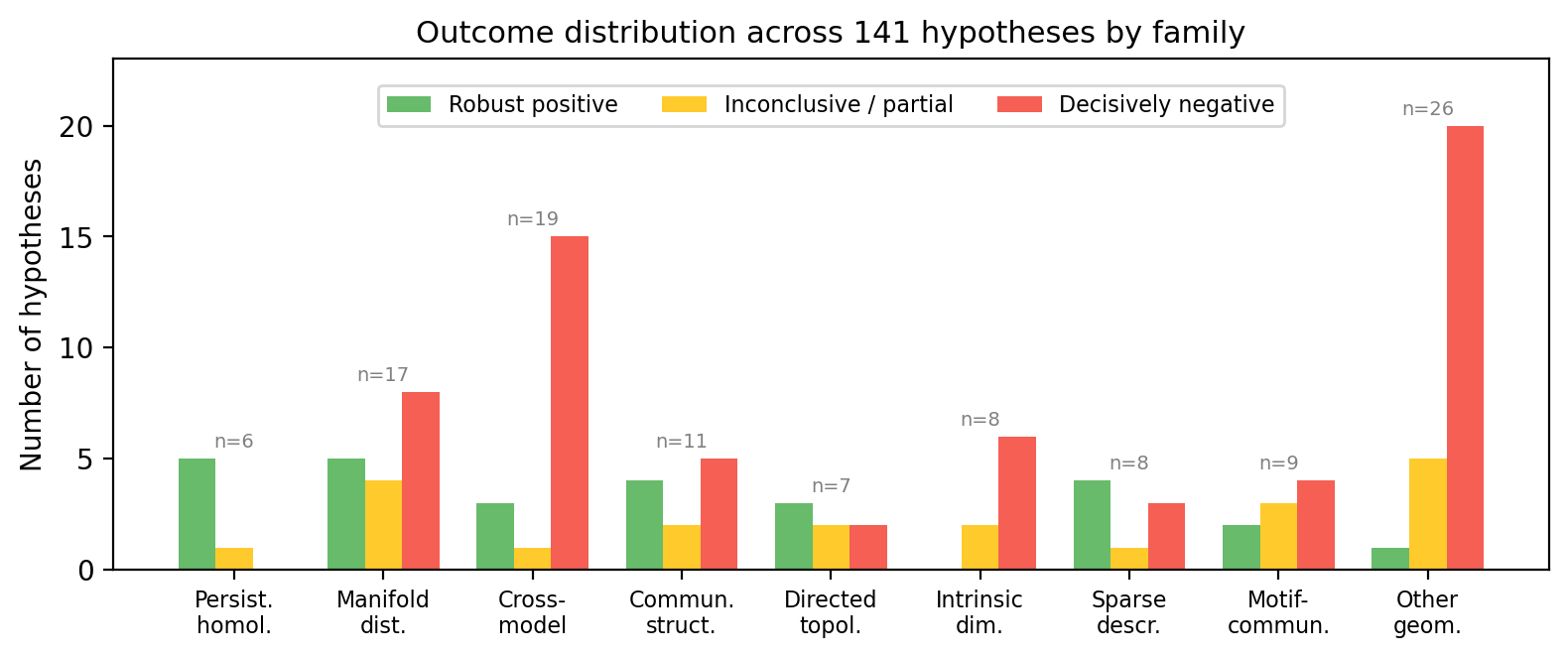}
\caption{Distribution of hypothesis outcomes across nine content families (111 of 141 total hypotheses; the remaining 30 are cross-cutting methodological variants---topology stability tests, null-framework development, split-design validation---that span multiple families). Approximately 27 showed positive results under their primary null control, 21 were inconclusive or partial, and 63 were decisively negative. Under the strictest max-null audit (Section~\ref{sec:strict_null}), fewer than 15 survive, concentrating the robust-positive rate to roughly 10\%.}
\label{fig:outcomes}
\end{figure*}

We organize findings by strength of evidence, presenting the most robust results first and building toward the critical calibration provided by strict null auditing.
Table~\ref{tab:summary} provides a summary; full iteration reports and raw results for all 141 hypotheses are available in the project repository.
Figure~\ref{fig:outcomes} shows the distribution of outcomes across all hypothesis families.

\begin{table*}[t]
\centering
\caption{Summary of the strongest positive and negative findings from 141 hypotheses. ``Null-gap'' indicates whether the signal survived above the 95th percentile of the relevant null distribution. Findings are ranked by strength of evidence.}
\label{tab:summary}
\small
\begin{tabular}{@{}p{4.5cm}p{2.5cm}p{2.0cm}p{2.0cm}p{4.0cm}@{}}
\toprule
Finding & Effect size & Null-gap & Verdict & Interpretation \\
\midrule
\multicolumn{5}{@{}l}{\textit{Strong positive findings}} \\
\addlinespace
Cross-model CCA alignment (H24) & $r = 0.80$ & $p < 10^{-5}$ & \textbf{Robust} & Models converge on similar gene-space geometry \\
Signed motif--community hardening (H123) & $\Delta$AUROC $+0.094$ & 22/22 rows & \textbf{Robust} & Regulatory motif signs align with community structure \\
Persistent homology (H01/H03) & $p < 0.006$ & 11/12 layers & \textbf{Robust} & Embedding neighborhoods contain non-trivial ``loops'' \\
Stability-selected descriptors (H91) & $\Delta$AUROC $+0.074$ & 6/6 splits & \textbf{Robust} & Multiple geometric features jointly predict regulation \\
Topology stability (H14/H44) & 12/12 layers & 3/3 domains & \textbf{Robust} & Topological signal survives bootstrap perturbations \\
Triangle-defect spectrum (H70) & $\Delta$AUROC $+0.026$ & 6/6 splits & \textbf{Robust} & Became backbone for all subsequent work \\
TRRUST sign-motif interaction (H116) & $\Delta$AUROC $+0.078$ & 6/6 splits & \textbf{Robust} & Breakthrough biology-anchored geometry finding \\
\addlinespace
\multicolumn{5}{@{}l}{\textit{Moderate positive findings}} \\
\addlinespace
Geodesic $>$ Euclidean distance (H13) & $\Delta$AUROC $+0.013$ & 7/12 layers & Moderate & Curved distances capture regulatory proximity \\
Bifiltration cycle-rank (H47) & $\Delta$AUROC $+0.006$ & 6/6 splits & Moderate & Independent topology method confirms H01/H03 \\
Directed/signed topology (H52) & $\Delta$AUROC $+0.015$ & 6/6 splits & Moderate & Directional information carries regulatory signal \\
Sectional anisotropy (H139) & $\Delta$AUROC $+0.031$ & 6/9 splits & Partial & Manifold curvature tracks regulation (mainly immune) \\
\addlinespace
\multicolumn{5}{@{}l}{\textit{Key negative findings}} \\
\addlinespace
Cross-model correspondence & Top-1 $< 0.01$ & --- & \textbf{Failed} & Models agree on shape but not gene placement \\
Rewiring-null topology (H07--H12) & 0/24 significant & --- & \textbf{Failed} & Topology intertwined with graph connectivity \\
Strict max-null audit (H141) & Margin $-0.005$ & 3/9 splits & \textbf{Fragile} & Robust signal concentrates in immune tissue \\
\bottomrule
\end{tabular}
\end{table*}

\subsection{Two Independently Trained Models Converge on the Same Geometric Organization}
\label{sec:cross_model}

Perhaps the most striking evidence that the geometric structure is biologically real---rather than an artifact of any single model's training---comes from cross-model comparisons.
scGPT and Geneformer were trained independently, on different datasets, with different architectures and objectives.
If both models converge on similar geometric organization of gene representations, that organization almost certainly reflects real biology.

We tested this with progressively refined alignment methods.
Canonical correlation analysis~\citep{hardoon2004canonical} with whitening (H24)---which projects both models' PCA-reduced gene embeddings into a shared subspace that maximizes inter-model correlation, then evaluates whether geometric relationships are preserved---revealed remarkable geometric consistency between matched scGPT and Geneformer gene sets: mean canonical correlation of 0.80, pairwise-distance Spearman correlation of 0.75, and gene-level top-1 retrieval accuracy of 72\% (combined Fisher $p = 3.17 \times 10^{-5}$ across all three tissue domains).
Simpler Procrustes alignment (H20) confirmed the pattern: 40\% top-1 retrieval accuracy, significant in all domains ($p = 1.6 \times 10^{-5}$).
Even the ranking of which geometric features are most informative was conserved across models (H17): centered cosine was the top-ranked feature in all three domains for both models ($p = 0.037$).

\begin{figure}[t]
\centering
\includegraphics[width=\columnwidth]{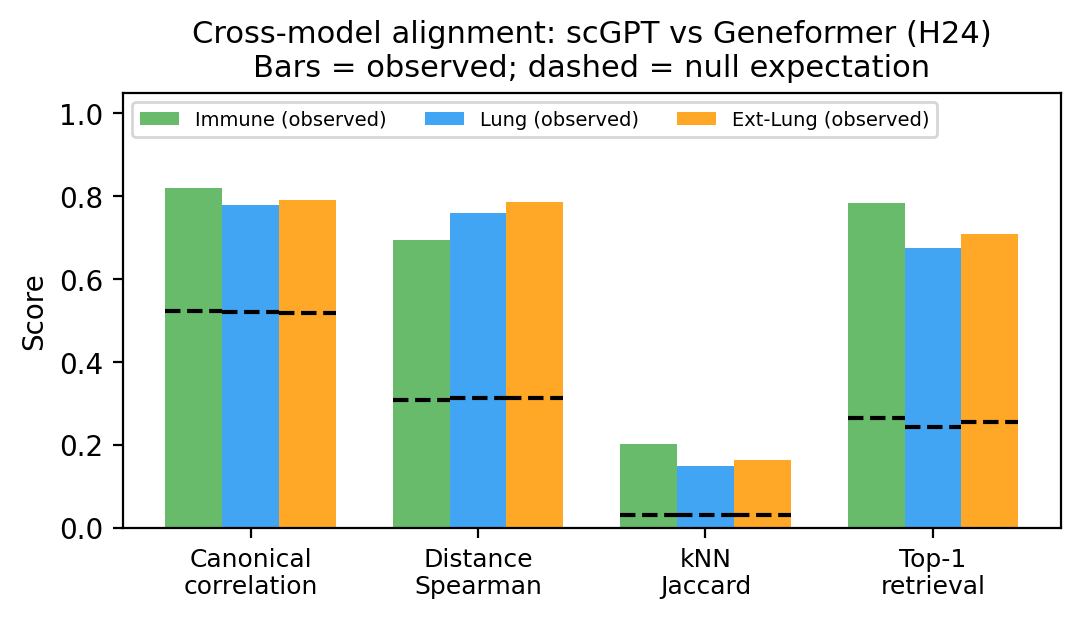}
\caption{Cross-model alignment between scGPT and Geneformer (H24). Bars show observed metrics; dashed lines show null expectations. Across all four alignment metrics and all three tissue domains, observed values substantially exceed null baselines, confirming that the two models converge on similar geometric organization despite independent training.}
\label{fig:cross_model}
\end{figure}

Two models trained on different data with different architectures arrive at the same geometric ``map'' of gene relationships.
This is analogous to independently constructed maps of a city agreeing on the positions of landmarks---strong evidence that the landmarks (biological relationships) are real features of the territory, not artifacts of the cartographer.

Despite this strong coarse agreement, recovering \emph{gene-level} correspondences between models proved nearly impossible.
Of 19 methods tested---including Gromov--Wasserstein transport, optimal transport, topology-signature distillation, cycle-consistent mapping, codebook transport, and chart/sheaf alignment---none achieved top-1 retrieval above 1\%.
The sole partial exception was perturbation-response rank alignment (H108), which achieved Spearman correlation of 0.73 and passed null-gap controls in 2/3 tissue domains---but failed in the immune domain ($-0.21$ null-gap) and did not replicate at multiseed resolution (H109: 2/9 rows passing).
The models agree on the ``shape'' of gene space---which genes are close together, which are far apart, which form clusters---but assign different internal coordinates to individual genes within that shared shape.
This distinction between coarse geometric agreement and fine-grained coordinate disagreement has practical implications: one cannot simply ``translate'' between models at the gene level, but geometric properties (distances, neighborhoods, clusters) are transferable.

\subsection{Gene Embedding Neighborhoods Contain Non-Trivial Topology}
\label{sec:persistent_homology}

We next asked whether scGPT's gene representations contain topological structure---specifically, whether the embedding neighborhoods form ``loops'' beyond what would be expected from random geometric arrangements.
Persistent homology~\citep{edelsbrunner2008persistent,zomorodian2005computing} detects such loops by constructing a filtration of simplicial complexes over increasing distance thresholds and identifying one-dimensional topological features (``H1 classes'') that persist across a range of thresholds; the total persistence (sum of birth-to-death lifetimes across all H1 classes) quantifies how much loop-like structure is present.

The answer is clearly yes, under appropriate controls.
Computing H1 persistence (via the Ripser algorithm~\citep{bauer2021ripser}) on PCA-projected embedding neighborhoods (350 genes, PCA to 20 dimensions) and comparing against feature-shuffle nulls (20 replicates per condition), we found that scGPT representations in the lung domain have significantly more topological structure than shuffled embeddings in 11 of 12 transformer layers ($p < 0.01$ in 9/12; top layer L0: $z = 3.21$, $p = 0.006$).
This signal replicated across all three tissue domains: immune (12/12 layers significant, mean H1 persistence delta $+12.1$ lifetime units) and external-lung (12/12 layers significant, mean delta $+12.5$).

\begin{figure*}[t]
\centering
\includegraphics[width=0.85\textwidth]{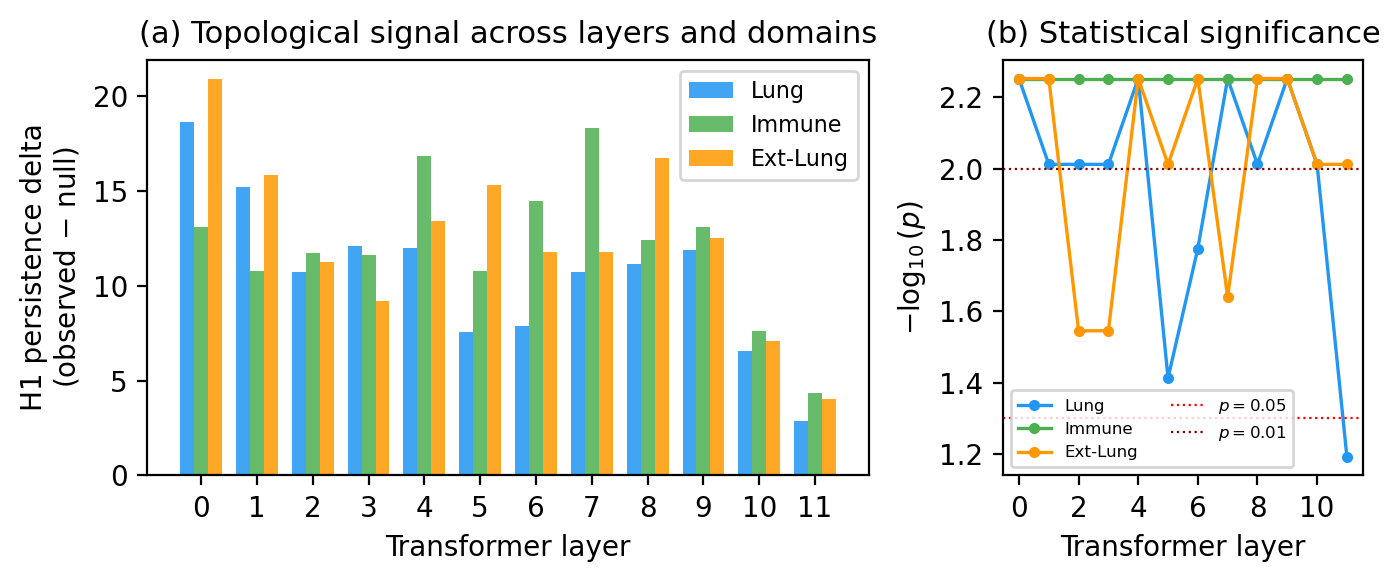}
\caption{Persistent homology across transformer layers and tissue domains. (a)~H1 persistence delta (observed minus null) for each layer, showing that topological structure exceeds null expectations across all domains and most layers, with characteristic peaks in early and middle layers. (b)~Statistical significance ($-\log_{10} p$) of the H1 signal; dashed lines mark $p = 0.05$ and $p = 0.01$ thresholds. The immune and external-lung domains show consistently strong significance, while the lung domain is more variable.}
\label{fig:persistence}
\end{figure*}

Crucially, the topology was stable: bootstrap testing across different sample sizes and PCA dimensions (10--20 components, $k \in \{8, 12, 16\}$) showed positive signal in 12/12 layers with complete parameter-setting stability, and true zigzag persistence---which tracks topological features that persist when alternating between two disjoint gene pools, testing whether loops are a property of the shared geometry rather than any particular gene subset---exceeded null expectations in all tested configurations.
A complementary topological measure---bifiltration-like cycle-rank (H47/H49), which counts the number of independent cycles in the $k$NN graph at each filtration threshold and uses this count as a per-edge feature for regulatory classification---also showed robust positive signal ($\Delta$AUROC $+0.006$, Fisher-significant in 6/6 domain-splits, 13/24 individual rows with $p < 0.05$), confirming that the topology is detectable through independent methods.

The ``loops'' detected by persistent homology suggest that genes are arranged in the embedding space along cyclic rather than purely hierarchical structures.
In biological regulatory networks, such cycles arise naturally from feedback loops (A activates B, B activates C, C inhibits A) and from the modular organization of pathways where genes at the ``boundaries'' between modules connect back to genes in the originating module.
The fact that this cyclic structure appears in all layers and domains suggests it reflects a fundamental property of how the model organizes genes.

Under degree-preserving $k$NN rewiring nulls---which shuffle the neighborhood graph while preserving how many neighbors each gene has---the topological signal vanished completely (0/24 layer-tests significant across six iterations of progressively refined rewiring controls).
This means the observed topology is not a deeper geometric invariant; rather, it arises from the specific pattern of which genes are each other's nearest neighbors.
The topology is \emph{real} in the sense that the model creates it (it's absent in shuffled embeddings), but it is \emph{fragile} in the sense that it depends on fine-grained neighborhood structure rather than robust global geometry.

\subsection{The Embedding Manifold Encodes a Distance Hierarchy for Regulation}
\label{sec:manifold_distance}

If the embedding space is geometrically structured, different ways of measuring inter-gene distance should capture different aspects of biological relationships.
We systematically tested multiple distance metrics and geometric features for their ability to discriminate regulatory gene pairs from non-regulatory pairs.

A clear hierarchy emerged.
Euclidean distance (straight-line in embedding space) was outperformed by geodesic distance (shortest path along the $k$NN manifold, with $k$ adaptively chosen from 10--35 to ensure graph connectivity; $\Delta$AUROC $+0.013$, 7/12 layers significant in both gene-pool splits).
Geodesic distance was in turn outperformed by diffusion distance~\citep{coifman2006diffusion}: the $L_2$ distance between rows of a diffusion-map embedding derived from a random walk on the $k$NN graph, with the diffusion time $t$ swept over $\{1, 2, 4, 8\}$ and selected by best performance ($\Delta$AUROC $+0.017$, significant in all three domains; Fisher $p$-values from $10^{-5}$ to $10^{-11}$).
Convexity-deficit and detour-based graph-geometry metrics provided additional evidence for the distance hierarchy (H32: $\Delta$AUROC $+0.017$, Fisher-significant in 4/6 domain-splits).
The strongest single geometry metric was the multiscale triangle-defect spectrum (H69/H70).
For each gene triplet in a $k$NN neighborhood, the triangle defect measures how much the pairwise distances deviate from the triangle inequality---quantifying local departure from flat Euclidean geometry.
Aggregating these defects at multiple neighborhood scales ($k \in \{8, 12, 16\}$) produces a per-edge feature vector that captures multiscale curvature around each gene pair.
This metric achieved $\Delta$AUROC $+0.026$ with positive null-gap in all domain-split groups under hard-null calibration.
This metric became the foundation for all subsequent work in the campaign: nearly every hypothesis from H70 onward was evaluated as an incremental improvement over this triangle-defect baseline.

\begin{figure*}[t]
\centering
\includegraphics[width=0.85\textwidth]{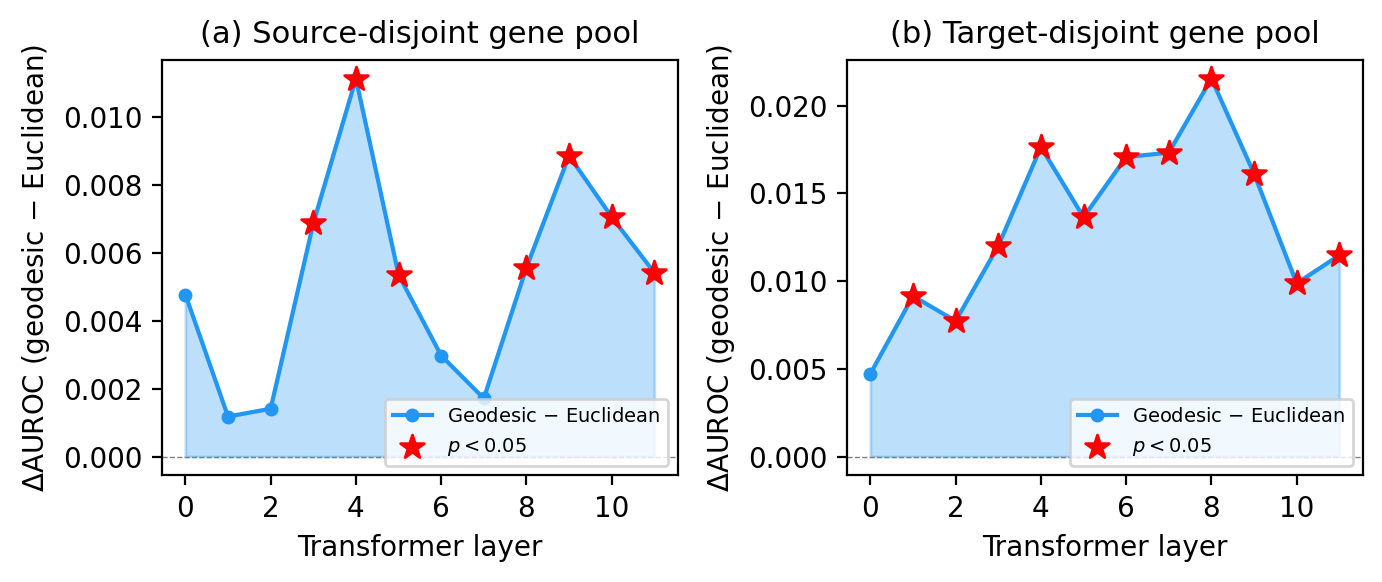}
\caption{Geodesic manifold distances outperform Euclidean for regulatory edge discrimination (H13). (a)~Source-disjoint and (b)~target-disjoint gene pool splits. The shaded area shows $\Delta$AUROC (geodesic minus Euclidean); red stars mark layers where the improvement is statistically significant ($p < 0.05$). The advantage is modest ($\Delta$AUROC $\approx 0.01$) but consistent across splits and concentrated in middle transformer layers.}
\label{fig:distance}
\end{figure*}

Regulatory gene pairs are not simply ``close'' in embedding space; they are connected by curved manifold paths that reflect the nonlinear way the model organizes biological relationships.
The diffusion distance advantage means that regulatory proximity is better captured by how easily a random walk can travel between genes along the manifold than by straight-line distance through high-dimensional space.
The triangle-defect spectrum further reveals that the local curvature of the manifold around regulatory gene pairs systematically differs from the curvature around non-regulatory pairs.

When diffusion distances were compared against a stricter coexpression-matched null, the uplift shrank from $+0.017$ to $+0.008$ and lost statistical significance in 2/3 domains.
This indicates that a substantial fraction of the diffusion distance advantage reflects the well-known tendency for coexpressed genes to be nearby in embedding space, rather than independent geometric information about regulation per se.
The triangle-defect spectrum, which survived matched-random-third-node controls, appears to capture truly independent geometric structure.

\subsection{The Strongest Finding: Regulatory Motifs Align with Geometric Community Structure}
\label{sec:motif_community}

The single most robust finding across the entire 141-hypothesis screen emerged from combining two sources of information: the geometric community structure of the embedding space (Louvain communities~\citep{blondel2008fast} detected on the $k$NN graph of gene embeddings) and the signed regulatory motif annotations from the TRRUST database (which transcription factors activate or repress which targets).

At a basic level, genes that are known to be in the same regulatory relationship are more likely to land in the same geometric community (H16: AUROC 0.54, all 12 layers significant in both gene-pool splits).
But the key finding is that the \emph{signs} of regulatory relationships (activation vs.\ repression) align with community structure in a way that is highly robust to null controls.

The breakthrough came with H116, which tested whether features derived from TRRUST signed regulatory annotations---specifically, whether a gene pair shares a common transcription factor with a known activation or repression sign, and whether the geometric community placement of that pair is consistent with the annotated sign---could predict regulatory edges above the triangle-defect baseline.
The result was strongly positive: $\Delta$AUROC $+0.078$, null-gap positive in 6/6 domain-splits---the first hypothesis in the campaign to achieve full null-gap coverage.
Building on this, H123 refined the approach with stricter null controls (TF-identity-preserving sign shuffles and motif-decoy shuffles matched on TF/target degree strata), and the signal strengthened further: positive in \textbf{every single test row} (22/22) and surviving the null-gap criterion in \textbf{every domain-split group tested}---the only hypothesis in the entire campaign to achieve complete null-gap coverage under the most stringent controls.
The effect size was substantial: $\Delta$AUROC $+0.094$ above the already-strong triangle-defect baseline.

\begin{figure*}[t]
\centering
\includegraphics[width=0.85\textwidth]{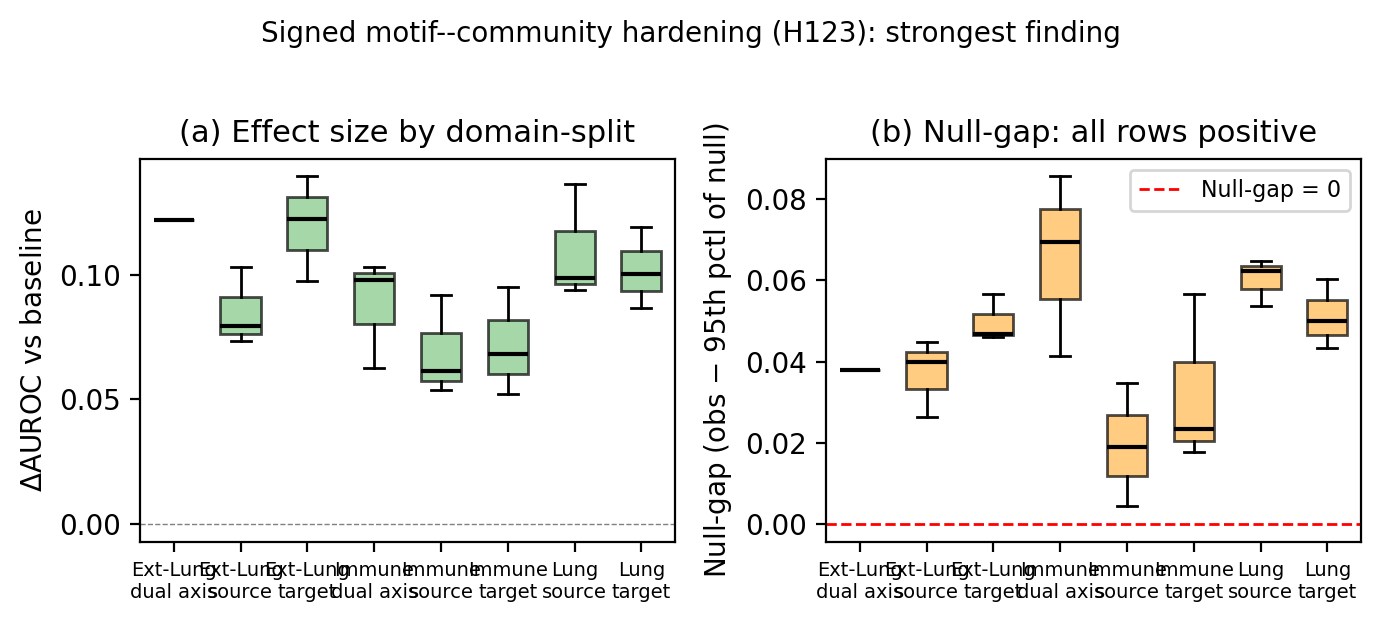}
\caption{Signed motif--community hardening (H123), the strongest finding across all 141 hypotheses. (a)~Effect size ($\Delta$AUROC vs.\ the H70 geometric baseline) across domain-split groups, showing consistently positive improvement. (b)~Null-gap analysis: the observed signal minus the 95th percentile of the null distribution is positive in \emph{all} test rows---the only hypothesis in the campaign to achieve complete null-gap coverage.}
\label{fig:motif}
\end{figure*}

The model doesn't just place regulated genes near their regulators; it organizes them so that activation targets and repression targets occupy geometrically distinguishable positions relative to the transcription factor within the community.
This is consistent with the model having learned something about the \emph{direction} and \emph{sign} of regulatory relationships, not just their existence.

Paradoxically, adding additional biological information consistently increased the raw effect size but \emph{systematically eroded} null-gap robustness.
The degradation followed a strikingly monotonic pattern: H123 alone (all domain-splits passing) $\to$ with STRING conditioning, H124 (fewer than half passing) $\to$ with GO co-membership, H127 (2/9 passing) $\to$ with continuous semantic GO similarity, H130 (0/9 passing) $\to$ with ontology sheaf features, H138 (0/9 passing).
Each additional annotation layer increased the raw $\Delta$AUROC (from $+0.094$ to $+0.134$) while progressively destroying robustness, until no domain-split survived strict null controls at all.
This counterintuitive pattern arose because the additional biological features were partially correlated with the null control structure, making it easier for null models to explain the signal.
The lesson: in interpretability research, more biological priors are not always better; they can introduce confounds that make it harder to isolate model-internal structure.

\subsection{Combining Geometric Features Through Stability Selection}
\label{sec:sparse_descriptors}

No single geometric measure captured all the regulatory information present in the embeddings.
This motivated a multivariate approach: for each gene pair, we computed a vector of geometric edge features (geodesic distance, triangle defects, community co-membership, bifiltration cycle-rank, directed topology features) and combined them via stability selection~\citep{meinshausen2010stability}---a procedure that fits randomized LASSO models on many bootstrap subsamples and retains only features selected in a high fraction of runs, yielding a sparse, reproducible feature set that is then used in a cross-validated logistic classifier.

The resulting stability-selected descriptors (H91) achieved $\Delta$AUROC $+0.074$, were positive in 72/72 test rows (three seeds, three domains, two splits, four layers), and showed positive null-gap in all 6/6 domain-split groups.
Critically, the descriptor selection was highly reproducible: the Jaccard similarity of selected features across different random seeds was 0.65, with sign agreement of 1.0.
A biologically anchored variant that incorporated confidence and regulatory-sign weighting (H93) achieved $\Delta$AUROC $+0.084$ with similarly complete null-gap coverage.

\begin{figure*}[t]
\centering
\includegraphics[width=0.75\textwidth]{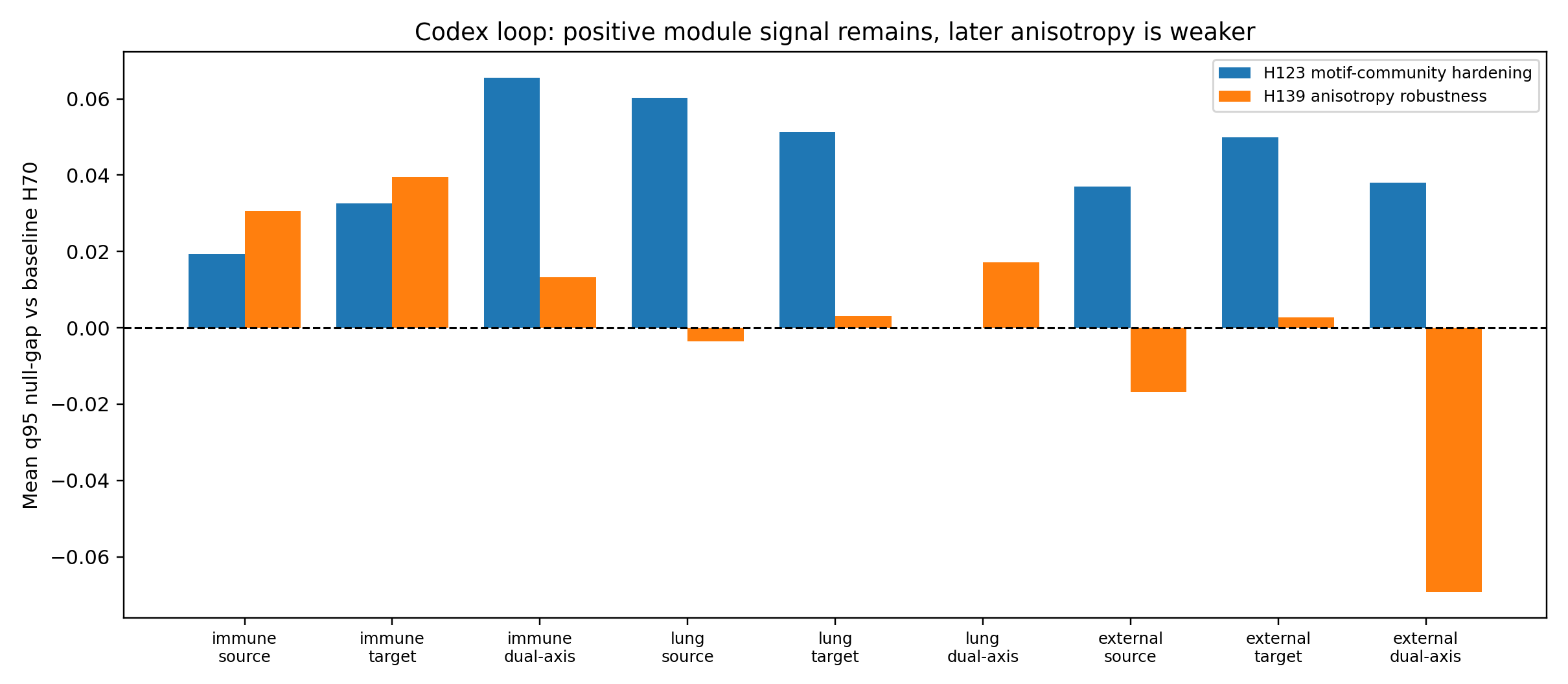}
\caption{Comparison of null-gap signal between the signed motif--community hardening (H123) and sectional anisotropy (H139) across domain-split groups, illustrating the sharp boundary between robust and fragile geometric findings.}
\label{fig:null_gap_comparison}
\end{figure*}

Biological regulatory information is distributed across multiple geometric dimensions of the embedding space---distance, topology, community structure, directionality---and no single dimension captures everything.
The consistency of feature selection across seeds indicates that this is a stable property of the representations, not an artifact of random initialization.

\subsection{A Sobering Calibration: The Strict Max-Null Audit}
\label{sec:strict_null}

All results above were evaluated against individual null families.
But how do they hold up under the most conservative possible test: comparing the observed signal against the \emph{maximum} of the 95th percentile across \emph{all} null families simultaneously?

\begin{figure*}[t]
\centering
\includegraphics[width=0.75\textwidth]{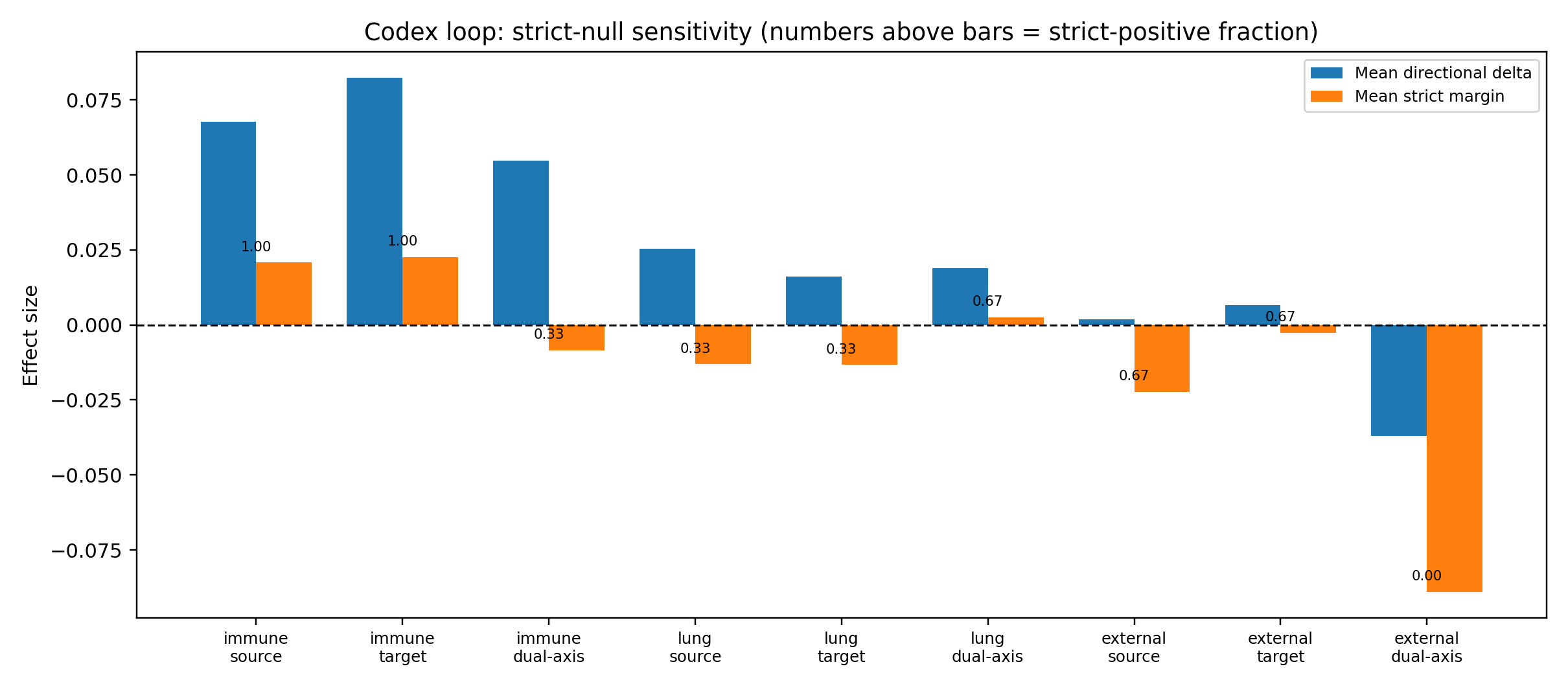}
\caption{Strict max-null fragility audit (H141). Each bar shows the strict margin (observed signal minus the maximum 95th percentile across all null families) for a domain-split group. Positive values (above the dashed line) indicate robust signal. The immune domain consistently survives strict auditing, while lung and external-lung are marginal or negative.}
\label{fig:strict_null}
\end{figure*}

The answer is sobering (Figure~\ref{fig:strict_null}).
The overall mean strict margin was $-0.005$, meaning that on average, the strongest null family could explain the observed signal.
Strict-positive support was found in only 3/9 domain-split groups (15/25 individual test rows).

At the domain level, a clear pattern emerged:
\begin{itemize}
\item \textbf{Immune}: positive strict margin ($+0.012$)---the signal is real and survives all controls.
\item \textbf{Lung}: slightly negative ($-0.008$)---the signal is marginal and may not survive the most stringent controls.
\item \textbf{External-lung}: negative ($-0.023$)---the signal is fragile under strict auditing.
\end{itemize}

\begin{figure*}[t]
\centering
\includegraphics[width=0.75\textwidth]{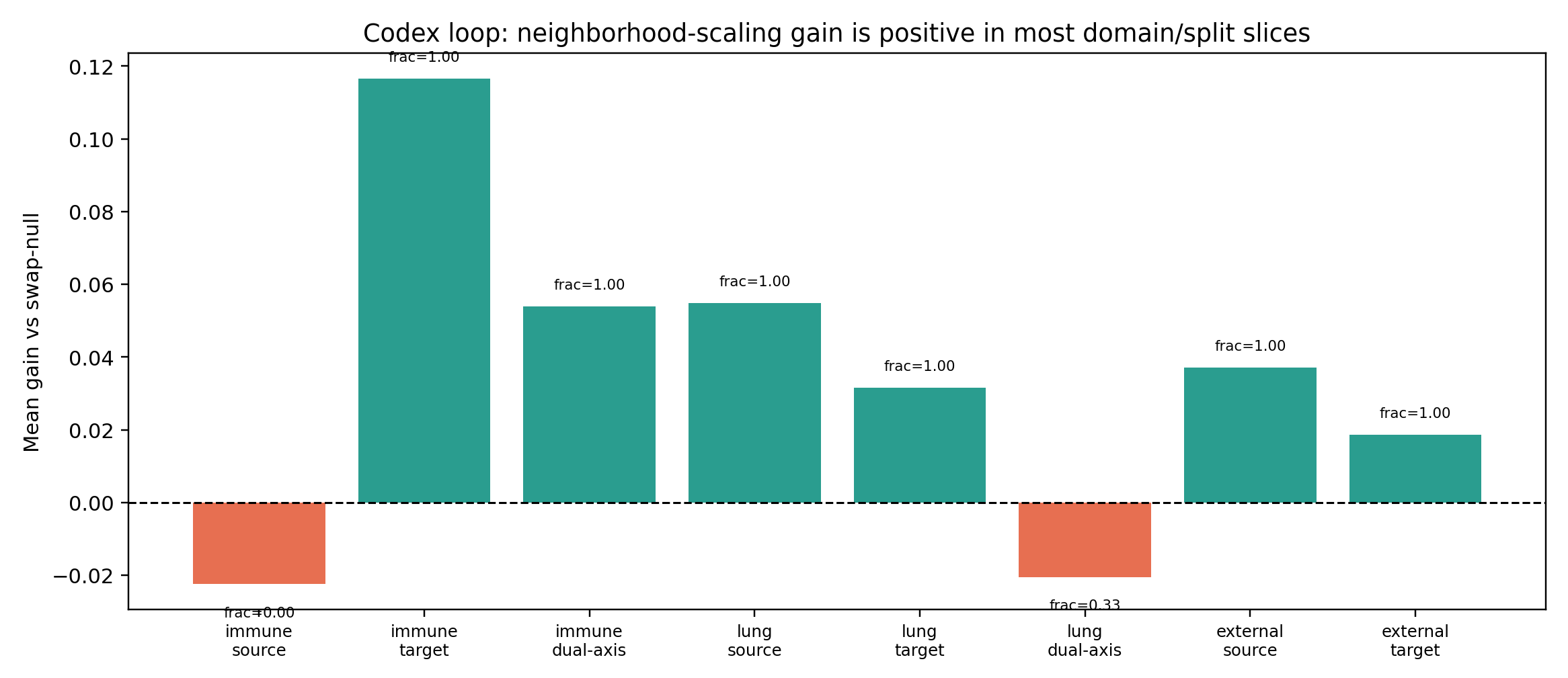}
\caption{Neighborhood scaling diagnostic (H140). Scaling the $k$NN neighborhood radius reveals how geometric signal varies with local resolution, providing a complementary view of where the embedding manifold encodes regulatory information.}
\label{fig:scaling}
\end{figure*}

The geometric structure we identified is genuine---the strongest signals (motif--community alignment, cross-model consistency, stability-selected descriptors) exceed every individual null family tested against them.
But it is \emph{localized}: concentrated in immune tissue and at the boundary of detection when all null families compete simultaneously.
This immune concentration was not unique to H141---it appeared as a recurring pattern throughout the campaign.
Diffusion distance uplift after covariate adjustment was immune-only (H31: immune Fisher $p = 6.8 \times 10^{-8}$; lung and external-lung non-significant).
Cross-model motif overlap was immune-only (H48: immune $p = 3.1 \times 10^{-4}$; lung and external-lung null).
Biologically anchored contrastive alignment was immune-only (H62: immune $p = 0.001$; other domains not robust).
The strict audit simply made this long-running pattern impossible to ignore.

This is consistent with two non-exclusive explanations.
First, the immune system has an unusually modular regulatory architecture (distinct T-cell, B-cell, and myeloid programs) that may create stronger geometric signatures than the less discretely organized regulatory programs in lung tissue.
Second, immune regulatory networks are better annotated in the databases we use as ground truth, so the ``failure'' in lung and external-lung may partly reflect annotation incompleteness rather than absence of geometric structure.

The practical implication is that claims about geometric structure in biological foundation models should be qualified by tissue context and null model hierarchy---a nuance often absent in the interpretability literature.

\subsection{What the Models Do Not Encode: 70+ Negative Results}
\label{sec:negative}

Among the most valuable outputs of this screening campaign are the hypotheses that were \emph{decisively rejected} (Table~\ref{tab:negative}).
Each negative result constrains interpretation and prevents overconfident claims about model geometry.

\begin{table}[t]
\centering
\caption{Key negative findings and their interpretive implications.}
\label{tab:negative}
\small
\begin{tabular}{@{}p{2.5cm}p{4.7cm}@{}}
\toprule
Failed hypothesis & What this tells us \\
\midrule
Topology robust to graph rewiring (H07--H12) & Topological ``loops'' vanish under degree-preserving rewiring; they depend on specific neighborhoods, not deep geometric invariants \\
\addlinespace
Forman curvature~\citep{sreejith2016forman} enrichment (H23) & High-curvature edges are \emph{less} likely to be regulatory (AUROC 0.34--0.39, below chance)---opposite to the hypothesis \\
\addlinespace
Cross-model correspondence (19 methods) & Models agree on geometry but not gene-level coordinates; ``translation'' between models is impossible at gene resolution \\
\addlinespace
Confidence-tier scaling (H19, H37) & Community structure tracks binary regulatory identity but not regulatory strength; the model knows \emph{whether} genes interact, not \emph{how strongly} \\
\addlinespace
Intrinsic dimension transfer (H42, H45) & In-sample manifold complexity correlates do not predict regulation out-of-sample ($\Delta R^2 = -10.7$); likely overfitting \\
\addlinespace
Hyperbolicity (H30) & The embedding manifold is not hyperbolic; tree-like hierarchical structure is not the right geometric metaphor for these representations \\
\addlinespace
Biological annotation extensions (H94, H127--H130, H135, H138) & Adding GO/STRING annotations inflates effect sizes but systematically degrades null-gap robustness \\
\bottomrule
\end{tabular}
\end{table}

Two patterns in the negative results deserve special attention.

At least 15 hypotheses showed large positive raw effects ($\Delta$AUROC $>0.03$) that vanished under strict null controls.
The most dramatic examples include the bridge-curvature lineage (H95/H97: 24/24 rows positive with mean $\Delta$AUROC $+0.079$, but 0/6 domain-splits surviving null-gap auditing) and the biologically anchored finite-state grammar (H111: $\Delta$AUROC $+0.112$, positive in all 6 domain-splits, but only 1/6 surviving null-gap---despite being one of the largest effect sizes observed across the entire campaign).
This pattern---``it looks great until you control properly''---is a cautionary tale for the interpretability field.
Without the hierarchical null framework, these results would have been reported as strong positive findings.

Across 19 distinct methods---Gromov--Wasserstein transport~\citep{peyre2019computational}, seeded and unseeded optimal transport, topology-signature distillation, cycle-consistent mapping, codebook transport, chart/sheaf alignment, and more---the mean top-1 retrieval rate for gene-level correspondence was below 1\%.
Only perturbation-response alignment (H108) showed partial promise, but it failed to replicate across seeds (H109).
This is not a failure of methodology; it is an intrinsic property of these models.
They learn similar \emph{relationships} between genes (distances, neighborhoods, clusters) but embed individual genes at different coordinates within that shared relational structure.

\section{Discussion}

\subsection{A Layered Picture of Geometric Structure}

This screening campaign reveals a layered picture of what biological foundation models encode geometrically, organized from the most robust to the most fragile:

\textbf{Layer 1: Coarse geometric consistency across models} (most robust).
scGPT and Geneformer agree on the global geometric organization of gene space---distances, neighborhoods, feature rankings---despite being trained independently.
This is the strongest evidence that the geometry reflects biology rather than model-specific artifacts.

\textbf{Layer 2: Non-trivial topology and community structure} (robust under feature-shuffle controls).
The embedding space contains non-trivial topological features (persistent loops) and community structure that tracks regulatory biology.
But these features are intertwined with the specific neighborhood graph rather than being deeper geometric invariants.

\textbf{Layer 3: Manifold distance hierarchy and directed geometry} (moderate).
The curved geometry of the embedding manifold carries regulatory information beyond straight-line distance, and directional features add signal beyond undirected geometry.
But the increments are modest ($\Delta$AUROC 0.01--0.03) and partially confounded with coexpression.

\textbf{Layer 4: Signed motif--community interactions} (strongest single signal, but requires external annotations).
Combining model geometry with TRRUST regulatory annotations yields the strongest null-robust finding, but requires knowledge of transcription factor identity and regulatory sign---information the model provides indirectly, not explicitly.

\textbf{Layer 5: Localized robustness under strict auditing} (fragile beyond immune tissue).
Under the most conservative null controls, robust signal concentrates in immune tissue, consistent with the immune system's distinctively modular regulatory architecture.

\subsection{Implications for Biological Model Interpretability}

These findings have practical implications for the growing field of biological model interpretability.

\textbf{Geometric structure is real but domain-dependent.}
Researchers should not assume that geometric interpretability findings generalize across tissues.
The same model (scGPT) showed robust geometric regulatory signal in immune tissue but fragile signal in lung tissue---likely reflecting differences in both the underlying regulatory architecture and the completeness of available annotations.

\textbf{Null model choice is paramount.}
Our 141-hypothesis screen demonstrated that the same geometric feature can appear highly significant under feature-shuffle controls but completely non-significant under rewiring controls or strict max-null auditing.
Any interpretability claim about biological model geometry should specify which null model it was tested against and acknowledge what would happen under stricter controls.

\textbf{Multivariate approaches outperform single metrics.}
No single geometric measure captured all available regulatory information.
Stability-selected combinations of multiple geometric features (distances, topology, community membership, directionality) substantially outperformed any individual measure, suggesting that biological regulatory information is distributed across multiple geometric dimensions of the embedding space.

\textbf{Cross-model consistency is the strongest test of biological reality.}
Features that are conserved across independently trained models provide much stronger evidence for biological meaning than features identified in a single model.
We recommend cross-model geometric testing as a standard validation step in biological interpretability work.

\subsection{The Value of Systematic Negative Results}

Of the 141 hypotheses tested, approximately 70 were decisively negative, 30 were inconclusive, and 40 showed at least directionally positive results.
Of these 40, roughly 27 survived their primary null control, but fewer than 15 survived the strict max-null audit (Section~\ref{sec:strict_null}).
This final ratio---roughly 10\% robust positive under the most conservative threshold---is itself informative.
It suggests that the space of ``geometric properties that could carry biological meaning'' is large, but the space of ``geometric properties that \emph{actually} carry biological meaning and survive rigorous controls'' is small and specific.

The systematic documentation of negative results prevents a common failure mode in interpretability research: the selective reporting of positive findings that creates an inflated picture of what models encode.
By publishing all 141 results---including the 70+ negatives---we provide a calibrated map of the boundary between real geometric structure and statistical artifact in biological foundation model representations.

\subsection{Limitations}

Our screening covered three tissue domains from a single species; extension to additional tissues, cell types, and species is needed to assess generality.
The regulatory ground truth databases (DoRothEA, TRRUST, STRING) are incomplete and biased toward well-studied gene families~\citep{schaefer2012tfcat,aibar2017scenic}, so some ``negative'' results may reflect annotation limitations.
The autonomous loop used a greedy exploration strategy; more systematic approaches (e.g., Bayesian optimization) might discover additional positive signals.
Finally, whether the geometric structure identified here translates to practical value for downstream biological tasks---perturbation prediction, drug target prioritization, pathway discovery---remains to be tested.

\section{Conclusion}

Through 141 hypotheses tested across 52 iterations of autonomous screening, we map the boundary between real geometric structure and artifact in biological foundation model representations.

The models do encode real structure: persistent topology, a manifold distance hierarchy, cross-model geometric consistency, community organization that tracks regulatory biology, and signed motif--community interactions that survive stringent null controls.
But this structure is more localized, more domain-dependent, and more fragile under strict auditing than optimistic interpretations would suggest.

The most important contribution may not be any single positive finding, but the comprehensive documentation of what \emph{doesn't} work: 70+ hypotheses that appeared promising but failed rigorous controls, 19 cross-model correspondence methods that consistently failed, and a strict max-null audit showing that robust signal concentrates in immune tissue.
These negatives, as much as the positives, define what we can and cannot extract from the geometry of biological foundation models.

\section*{Data and Code Availability}

All experimental scripts, iteration reports, raw CSV results, and the autonomous loop orchestration code are available in the project repository at \url{https://github.com/ihorkendiukhov/biomechinterp}.

\bibliographystyle{plainnat}

\end{document}